\pdfoutput=1
\documentclass[letterpaper,twocolumn,10pt]{article}
\usepackage{usenix,epsfig,endnotes}
\usepackage{url}
\usepackage{amsmath}
\usepackage{amssymb}
\usepackage{amsfonts}
\usepackage{caption} 
\usepackage{todonotes}
\usepackage{cite}
\usepackage{verbatim}
\usepackage{subfigure}
\usepackage{epstopdf}
\usepackage{capt-of}
\usepackage{floatrow}
\usepackage{fancyvrb}
\usepackage{lipsum}

\newfloatcommand{capbtabbox}{table}[][\FBwidth]

\newcommand\PnP{\texttt{Prime+Probe}}	
\title{CacheZoom: How SGX Amplifies The Power of Cache Attacks}

\author{
{\rm Ahmad Moghimi}\\
Worcester Polytechnic Institute\\
amoghimi@wpi.edu
\and
{\rm Gorka Irazoqui}\\
Worcester Polytechnic Institute\\
girazoki@wpi.edu
\and
{\rm Thomas Eisenbarth}\\
Worcester Polytechnic Institute\\
teisenbarth@wpi.edu
}

\begin{document}
\maketitle
\thispagestyle{empty}

\begin{abstract}
In modern computing environments, hardware resources are commonly shared, and parallel computation is widely used. Parallel tasks can cause privacy and security problems if proper isolation is not enforced. Intel proposed SGX to create a trusted execution environment within the processor. SGX relies on the hardware, and claims runtime protection even if the OS and other software components are malicious. However, SGX disregards side-channel attacks. We introduce a powerful cache side-channel attack that provides system adversaries a high resolution channel. Our attack tool named \emph{CacheZoom} is able to virtually track all memory accesses of SGX enclaves with high spatial and temporal precision. As proof of concept, we demonstrate AES key recovery attacks on commonly used implementations including those that were believed to be resistant in previous scenarios. Our results show that SGX cannot protect critical data sensitive computations, and efficient AES key recovery is possible in a practical environment. In contrast to previous works which require hundreds of measurements, this is the first cache side-channel attack on a real system that can recover AES keys with a minimal number of measurements. We can successfully recover AES keys from T-Table based implementations with as few as ten measurements.\footnote{
\noindent\textbf{Publication} This work is accepted at the Conference on Cryptographic Hardware and Embedded Systems (CHES 2017)~\cite{cryptoeprint:2017:618}.
}
\end{abstract}

\section{Motivation}

In the parallel computing environment, processes with various trust and criticality levels are allowed to run concurrently and share system resources. Proliferation of cloud computing technology elevated these phenomena to the next level. Cloud computers running many different services authored by various
providers process user information on the same hardware. Traditionally, the operating system (OS) provides security and privacy services. In cloud computing, cloud providers and the hypervisor also become part of the Trusted Computing Base (TCB). Due to the high complexity and various attack surfaces in modern computing systems, keeping an entire system secure is usually unrealistic~\cite{langner2011stuxnet,dahbur2011survey}. 

One way to reduce the TCB is to outsource security-critical services to Secure Elements (SE), a separate trusted hardware which usually undergoes rigorous auditing. Trusted Platform Modules (TPM), for example, provide services such as cryptography, secure boot, sealing data and attestation beyond the authority of the OS \cite{morris2011trusted}. However, SEs come with their own drawbacks: they are static components and connected to the CPU over an untrusted bus. Trusted Execution Environments (TEE) are an alternative, which provide similar services within the CPU. A TEE is an isolated environment to run software with a higher trust level than the OS. The software running inside a TEE has full access to the system resources while it is protected from other applications and the OS. Examples include ARM TrustZone \cite{armTrust} and Intel Software Guard eXtensions (SGX) \cite{inteSGX}. Intel SGX creates a TEE on an untrusted system by only trusting the hardware in which the code is executed. Trusted code is secured in an \emph{enclave}, which is encrypted and authenticated by the hardware. The CPU decrypts and verifies enclave code and data as it is moved into the cache. That is, enclaves are logically protected from malicious applications, the OS, and physical adversaries monitoring system buses. However, Intel SGX is not protected against attacks that utilize hardware resources as a side channel~\cite{IntelTutorial}. And indeed, first proposed works showing that microarchitectural side channels can be exploited include attacks using page table faults~\cite{xu2015controlled} and the branch prediction unit~\cite{lee2017Infer}.

Caches have become a very popular side channel in many scenarios, including mobile~\cite{lipp2015armageddon} and cloud environments~\cite{Inci2016}. Reasons include that Last Level Cache (LCC) attacks perform well in cross-core scenarios on Intel machines. Another advantage of cache attacks are the high \emph{spatial} resolution they provide. This high spatial resolution, combined with a good temporal resolution, have enabled attacks on major asymmetric implementations, unless they are optimized for constant memory accesses. For symmetric cryptography, the scenario is more challenging. A software AES implementation can be executed in a few hundred cycles, while a \PnP\ cycle on the LLC takes about 2000 cycles to monitor a single set. To avoid the undersampling, synchronized attacks first prime, trigger a single encryption and then probe, yielding at best one observation per encryption~\cite{bernstein}. Higher resolution is only possible in OS adversarial scenarios.

\subsection{Our Contribution}
We demonstrate not only that Intel SGX is vulnerable to cache attacks, but show that with SGX, the quality of information retrieved is significantly improved. The improved resolution enables attacks that are infeasible in previous scenarios, e.g., cloud environments. We utilize all the capabilities that SGX assumes an attacker has, i.e., full access to OS resources. We  construct a tool\footnote{\emph{CacheZoom}  source and data sets: \url{https://github.com/vernamlab/CacheZoom}}
 named \emph{CacheZoom} that is able to interrupt the victim every few memory accesses, thereby collecting high-resolution information about all memory accesses that the target enclave makes by applying \PnP\ attack in the L1 cache. The usage of core-private resources does not reduce the applicability of the attack, as the compromised OS schedules both victim and attacker in the same core.

While tracking memory accesses of enclave with high temporal and spatial resolution has many adversarial scenarios, we demonstrate the power of this side channel by attacking several AES implementations. Further, we show that adopted countermeasures in popular cryptographic libraries, like cache prefetching and implementations with small memory footprint, not only do not prevent attacks, but can facilitate attacker's observation. In short, this work:

\begin{itemize}
\item Presents a powerful and low-noise side channel implemented through the L1 cache. We exploit several capabilities corresponding to the compromised OS. This side channel can be applied against TEEs to recover fine grained information about memory accesses, which often carry sensitive data.
\item Demonstrates the strength of our side channel by recovering AES keys with fewer traces than ever in previous attacks, and further, by attacking implementations considered resistant against cache attacks.
\item Shows that some of the countermeasures that were supposed to protect AES implementations, e.g. prefetching and S-box implementations, are not effective in the context of SGX. In fact, prefetching can even ease the retrieval of memory traces.
\end{itemize}

\section{Background}\label{sec:background}
This section covers topics that help understand the side channel used to retrieve sensitive information. We discuss the basic functionality of Intel SGX and possible microarchitectural attacks that can be deployed against it.

\subsection{How Intel SGX Works}
Intel introduced SGX, a new subset of hardware instructions that allows execution of software inside isolated environments called \emph{enclaves} with the release of Skylake generation. Enclaves are isolated from other components running on the same hardware, including OSs. SGX has recently gained attention of the security community and various SGX-based solutions have been proposed~\cite{arnautov2016scone, baumann2015shielding, schuster2015vc3}.

Enclave modules can be shipped as part of an untrusted application and can be utilized by untrusted components of the application. The untrusted component interacts with the system software, which dedicates specific trusted memory regions for the enclave. After that, the authenticity, integrity and confidentiality of enclave are provided and measured by the hardware. Any untrusted code base, including the OS, has no control over the trusted memory region. Untrusted applications can only use specific instructions to call the trusted component through predefined interfaces. This design helps developers to benefit from the hardware isolation for security critical applications. 

SGX is designed to protect enclaves from malicious users that gain root access to an OS. Memory pages belonging to an enclave are encrypted in DRAM and protected from a malicious OS snooping on them. Pages are only decrypted when they are processed by the CPU, e.g., when they are moved to the caches. In short, SGX assumes only the hardware to be trusted; any other agent is considered susceptible of being malicious. Upon enclave creation, virtual memory pages that can only map to a protected DRAM region (called the Enclave Page Cache) are reserved. The OS is in charge of the memory page mapping; however, SGX detects any malicious mapping performed by it. In fact, any malicious action from the OS will be stored by SGX and is verifiable by third party agents.

\subsection{Microarchitectural Attacks in SGX}
Despite all the protection that SGX offers, the documentation specifically claims that side channel attacks were not considered under the threat scope of its design. In fact, although dealing with encrypted memory pages, the cache utilization is performed similar to decrypted mode and concurrently to any other process in the system. This means that the hardware resources can be utilized as side channels by both malicious enclaves and OSs. While enclave-to-enclave attacks have several similarities to cross-VM attacks, malicious OS-to-enclave attacks can give attackers a new capability not observed before: virtually unlimited temporal resolution. The OS can interrupt the execution of enclave processes after every small number of memory accesses to check the hardware utilization, as just the TLB (but no other hardware resources) is flushed during context switches. Further, while cross-core attacks gained huge popularity in others scenarios for not requiring core co-residency, a compromised OS can assign an enclave any affinity of its choice, and therefore use any core-private resource. Thus, while SGX can prevent untrusted software to perform Direct Memory Access (DMA) attacks, it also gives almost full resolution for exploitation by hardware side channels. For instance, an attacker can exploit page faults to learn about the memory page usage of the enclave. Further she can create contention and snoop on the utilization of any core-private and core-shared resource, including but not limited to Branch Prediction Units (BPUs), L1 caches or LLCs~\cite{aciiccmez2007power,osvik2006cache,liu2015last}. Further, although applicable in other scenarios~\cite{Bhattacharya2015}, enclave execution mode does not update the Hardware Performance Counters, and these can not provide (at least directly) information about the isolated process. 

From the aforementioned resources, cache gives the most information. Unlike page faults, which at most will give a granularity of 4\,kB, cache hits/misses can give 64\,byte utilization granularity. In addition, while other hardware resources like Branch Prediction Units (BPU) can only extract branch dependent execution flow information, cache attacks can extract information from any memory access. Although most prior work targets the LLC for being shared across cores, this is not necessary in SGX scenarios, local caches are as applicable as LLC attacks. Further, because caches are not flushed when the enclave execution is interrupted, the OS can gain almost unlimited timing resolution.

\subsection{The Prime+Probe Attack}
The \PnP\ attack was first introduced as a spy process capable of attacking core-private caches~\cite{osvik2006cache}. It was later expanded to recover RSA keys~\cite{aciiccmez2008vulnerability}, keystrokes and ElGamal keys across VMs~\cite{Zhang:2012:CSC:2382196.2382230, ristenpart2009hey}. As our attack is carried out in the L1 caches, we do not face some hurdles (e.g. slices) that an attacker would have to overcome. The \PnP\ attack stages include:
\begin{itemize}
\item{\bf Prime:} in which the attacker fills the entire cache or a small portion of it with her own dummy data.
\item{\bf Victim Access:} in which the attacker waits for the victim to make accesses to particular sets in the cache, hoping to see key dependent cache utilization. Note that, in any case, victim accesses to primed sets will evict at least one of the attackers dummy blocks from the set.
\item{\bf Probe:} in which the attacker performs a per-set timed re-access of the previously primed data. If the attacker observes a high probe time, she deduces that the cache set was utilized by the victim. On the contrary, if the attacker observes low access times, she deduces that all the previously primed memory blocks still reside in the cache, i.e., it was not utilized by the victim.
\end{itemize}

Thus, the \PnP\ methodology allows an attacker to guess the cache sets utilized by the victim. This information can be used to mount a full key recovery attack if the algorithm has key-dependent memory accesses translated into different cache set accesses.

\section{Related Work}

\textbf{Timing side-channel attacks} have been studied for many years. On a local area network, the timing of the decryption operation on a web server could reveal information about private keys stored on the server \cite{brumley2005remote}. Timing attacks are capable of breaking important cryptography primitives, such as exponentiation and factorization operations of Diffie-Hellman and RSA \cite{kocher1996timing}. More specifically, \textbf{microarchitectural timing side channels} have been explored extensively \cite{ge2016survey}. The first attacks proposed were based on the timing difference between local core-private cache misses and hits. Generally, cache timing attacks are based on the fact that a spy process can measure the differences in memory access times. These attacks are proposed to recover cryptography keys of ciphers such as DES~\cite{tsunoo2003cryptanalysis}, AES~\cite{bonneau2006cache} and RSA~\cite{percival2005cache}. Although there exist solutions to make cryptographic implementation resistant to cache attacks \cite{brickell2006software, osvik2006cache}, most of these solutions result in worse performance. Further, cache attacks are capable of extracting information from non-cryptographic applications~\cite{Zhang:2014:CSA:2660267.2660356}.

More recent proposals applied \textbf{cache side channels on shared LLC}, a shared resource among all the cores. This is important as, compared to previous core-private attacks, LLC attacks are applicable even when attacker and victim reside in different cores. The Flush+Reload~\cite{yarom2014flush+, benger2014ooh} attack on LLC is only applicable to systems with shared memory. These side channels can be improved by performance degradation~\cite{gullasch2011cache, allan2016amplifying}. Flush+Reload can be applied across VMs~\cite{waitaminute}, in Platform as a service (PaaS) clouds~\cite{Zhang:2014:CSA:2660267.2660356} and on smartphones~\cite{lipp2015armageddon}. The Flush+Reload is constrained by the memory deduplication requirement. On the other hand, \PnP~\cite{liu2015last}, shows that in contrast to the previous core-private cache side channels and the Flush+Reload attack, practical attacks can be performed without memory deduplication or a core co-residency requirement. The \PnP\ attack can be implemented from virtually any cloud virtual machines running on the same hardware. The attacker can identify where a particular VM is located on the cloud infrastructure such as Amazon EC2, create VMs until a co-located one is found~\cite{ristenpart2009hey, zhang2011homealone} and perform cross-VM \PnP\ attacks~\cite{irazoqui2015s}. \PnP\ can also be mounted from a browser using JavaScript~\cite{oren2015spy} and as a malicious smartphone application~\cite{lipp2015armageddon}. In addition to caches, other microarchitectural components such as \textbf{Branch Target Buffers (BTB)} are vulnerable to side channels~\cite{aciiccmez2007power, lee2017Infer}. %BTB is a shared processor cache used to predict the target of a branch before its execution. 
BTB can be exploited to determine if a branch has been taken by a target process or not, e.g. to bypass Address Space Layout Randomization (ASLR)~\cite{evtyushkin2016jump}.

\textbf{Security of Intel SGX} has been analyzed based on the available public resources~\cite{costanintel}. A side channel resistant TCB is proposed in the literature~\cite{SanctumCostan}. However, the proposed solution requires significant changes to the design of the processor. Similar to Intel SGX, ARM TrustZone is vulnerable to cache side-channel attacks~\cite{lipp2015armageddon}. Control-Channel attacks~\cite{xu2015controlled} have been proposed using the page-fault mechanism. An adversarial OS can introduce page faults to a target application and, based on the timing of the accessed page, the execution flow of a target can be inferred at page size granularity. Page fault side channels are effective on SGX and can be defeated using software solutions~\cite{ShindeYourFault} or by exploiting Intel Transactional Synchronization Extensions (TSX)~\cite{Shih2017TSGX}. Race conditions between two running threads inside an enclave can be exploited~\cite{weichbrodt2016asyncshock}. SGX-Shield~\cite{Seo2017SGXShield} proposes protection by adding ASLR protection and introduces software diversity inside an enclave. 
Several Cache attacks on SGX have recently and concurrently been shown, e.g. on AES~\cite{gotzfried2017cache} and RSA~\cite{brasser2017software}. While those works also exploit core co-location and L1 cache leakage, they fall short of exposing the full temporal and spatial resolution and thus focus on known vulnerable implementations and attack styles.
An enclave-to-enclave attack through LLC in a different adversarial scenario~\cite{malwareguard}, as well as methods to detect privileged side-channel attacks from within an enclave~\cite{chen2017detecting} have concurrently been proposed.
%In concurrent work: Cache attacks on SGX running AES~\cite{gotzfried2017cache} and RSA~\cite{brasser2017software} with less temporal and spatial resolutions, enclave-to-enclave attack through LLC in a different advarsarial scenario~\cite{malwareguard}, and detecting privileged side-channel attacks~\cite{chen2017detecting} are proposed.

\section{Creating a High-resolution Side Channel on Intel SGX}
We explain how to establish a high resolution channel on a compromised OS to monitor an SGX enclave. We first describe attacker capabilities, then our main design goals and how our malicious kernel driver is implemented. We finally test the resolution of our proposed side channel.

\subsection{Attacker Capabilities}\label{AttackerCapabilities}
In our attack, we assume that the adversary has root access to a Linux OS running SGX. The attacker is capable of installing kernel modules and configuring boot properties of the machine. As consequence of root access, the attacker can read the content of static binary on the disk, observe which symmetric cipher and implementation is used, and \emph{identify offset of tables that static data from the victim binary will occupy}.\footnote{If the enclave binary is obfuscated, position of tables needs to be reconstructed using reverse engineering methods, e.g. by analyzing cache access patterns~\cite{Inci2016}.} Although the attacker can observe the binary, she has no knowledge of the cipher key used during the encryption. In addition, the attacker is capable of synchronizing the enclave execution with CacheZoom. These assumptions are reasonable, as SGX promises a trusted environment for execution on untrusted systems. Symmetric keys can be generated at runtime from a secure random entropy (using \emph{RDRAND} instruction) and/or transferred through a public secure channel without the attacker knowledge.

\subsection{CacheZoom Design}
To create a high bandwidth channel with minimal noise, \textbf{(1)} we need to isolate the attackers' malicious spy process code and the target enclave's trusted execution from the rest of the running operations and \textbf{(2)} we need to perform the attack on small units of execution. By having these two elements, even a comparably small cache like L1 turns into a high capacity channel. Note that our spy process monitors the L1D data cache, but can also be implemented to monitor the L1I instruction cache or LLC. Our spy process is designed to profile all the sets in the L1D cache with the goal of retrieving maximum leakage. In order to avoid noise, we dedicate one physical core to our experimental setup, i.e., to the attacker \PnP\ code and the victim enclave process. All other running operations on the system, including OS services and interrupts, run on the remaining cores. Furthermore, CacheZoom forces the enclave execution to be interrupted in short time intervals, in order to identify all enclave memory accesses. Note that, the longer the victim enclave runs without interruption, the higher the number of accesses made to the cache, implying higher noise and less temporal resolution. CacheZoom should further reduces other possible sources of noise, e.g., context switches. The main purpose is that the attacker can retrieve most of the secret dependent memory accesses made by the target enclave. Since the L1 cache is virtually addressed, knowing the offset with respect to a page boundary is enough to know the accessed set.

\subsection{CacheZoom Implementation}
We explain technical details behind the implementation of Cache\-Zoom, in particular, how the noise sources are limited and how we increase the time resolution to obtain clean traces.

\subsubsection{Enclave-Attack Process Isolation}
Linux OS schedules different tasks among available logical processors by default. The main scheduler function \path{__schedule} is triggered on every tick of the logical processor's local timer interrupt. One way to remove a specific logical processor from the default scheduling algorithm is through the kernel boot configuration \path{isolcpus} which accepts a list of logical cores to be excluded from scheduling. To avoid a logical core from triggering the scheduling algorithm on its local timer interrupt, we can use \path{nohz_full} boot configuration option. Recall that reconfiguring the boot parameters and restarting the OS is included in our attackers capabilities. However, these capabilities are not necessary, as we can walk through the kernel task tree structure and turn the \path{PF_NO_SETAFFINITY} flag off for all tasks. Then, by dynamically calling the kernel \path{sched_setaffinity} interface for every task, we are able to force all the running kernel and user tasks to execute on specific cores. In addition to tasks and kernel threads, interrupts also need to be isolated from the target core. Most of the interrupts can be restricted to specific cores except for the non-maskable interrupts (NMIs), which can't be avoided. However, in our experience, their occurrence is negligible and does not add considerable amount of noise.

CPU frequency has a more dynamic behavior in modern processors. Our target processor has \textbf{Speedstep} technology which allows dynamic adjustment of processor voltage and \textbf{C-state}, which allows different power management states. These features, in addition to hyper-threading (concurrent execution of two threads on the same physical core), make the actual measurement of cycles through \textit{rdtsc} less reliable. Cache side channel attacks that use this cycle counter are affected by the dynamic CPU frequency. In non-OS adversarial scenarios, these noise sources have been neglected thus forcing the attacker to do more measurements. In our scenario, these processor features can be disabled through the computer BIOS setup or can be configured by the OS to avoid unpredictable behavior. In our attack, we simply disable every second logical processor to practically avoid hyper-threading. To maintain a stable frequency in spite of the available battery saving and frequency features, we set the CPU scaling governor to \textbf{performance} and limit the maximum and minimum frequency range.

\subsubsection{Increasing the time resolution}
Aiming at reducing the number of memory accesses made by the victim between two malicious OS interrupts, we use the local APIC programmable interrupt, available on phyisical cores. The APIC timer has different programmable modes but we are only interested in the \textbf{TSC-Deadline} mode. In TSC deadline mode, the specified TSC value will cause the local APIC to generate a timer IRQ once the CPU reaches it. In the Linux kernel, the function \path{lapic_next_deadline} is responsible for setting the next deadline on each interrupt. The actual interrupt handler routine for this IRQ is \path{local_apic_timer_interrupt}. In order to enable/disable our attack, we install hooks on these two functions. By patching the null function calls, available for the purpose of live patching, we can redirect these functions to the malicious routines in our kernel modules at runtime.

{\small
\begin{verbatim}
ffffffff81050900 lapic_next_deadline
ffffffff81050900: callq  null_sub1
\end{verbatim}

\begin{verbatim}
ffffffff81050c90 local_apic_timer_interrupt
ffffffff81050c90:	callq  null_sub2
\end{verbatim}
}

In the modified \path{lapic_next_deadline} function, we set the timer interrupt to specific values such that the running target enclave is interrupted every short period of execution time.  In the modified \path{local_apic_timer_interrupt}, we first probe the entire 64 sets of the L1D cache to gather information of the previous execution unit and then prime the entire 64 sets for the next one. After each probe, we store the retrieved cache information to a separate buffer. Our kernel driver is capable of performing 50000 circular samplings. To avoid unnecessarly sampling, we need to synchronize with the target enclave execution. For this purpose, we enable the hooks just before the call to the enclave interface and disable it right after.

\begin{figure}[t!]%
  \centering
  \includegraphics[width=\linewidth]{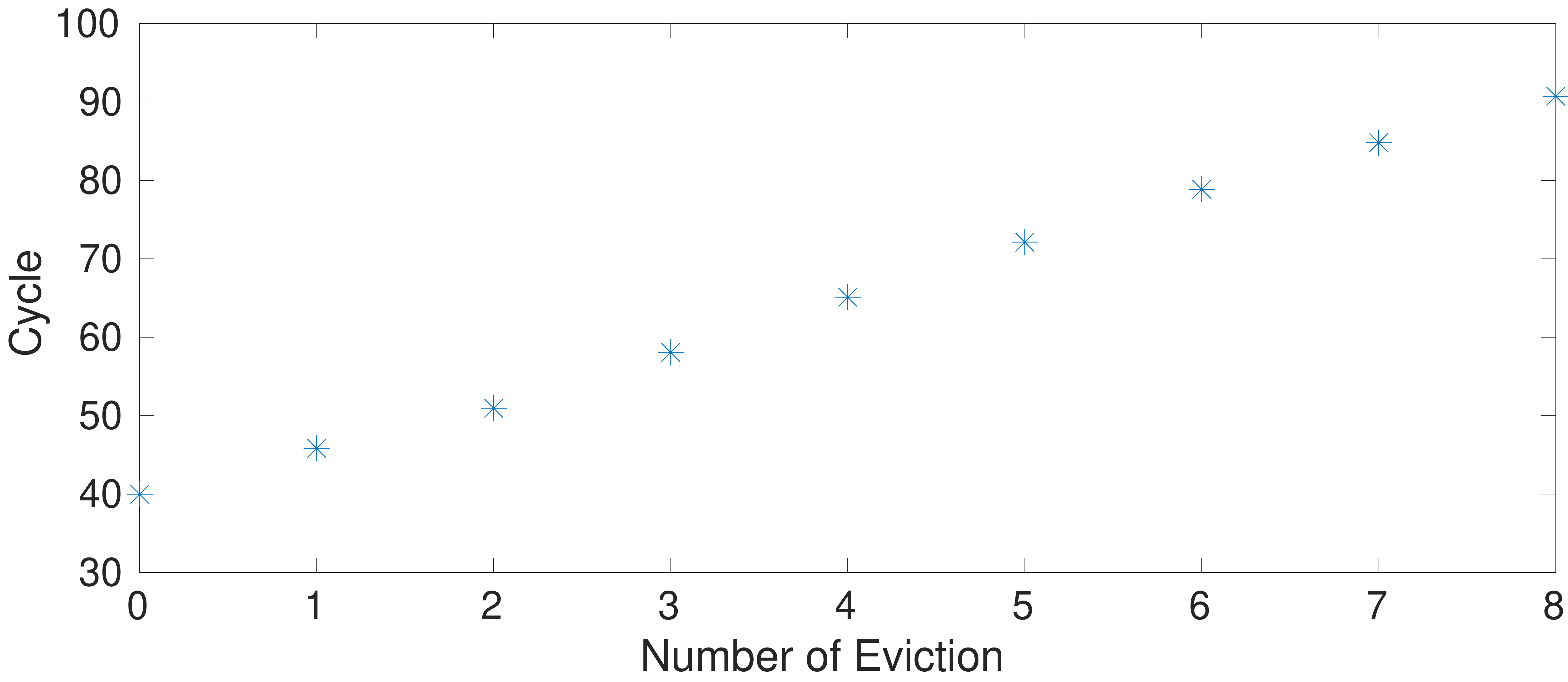}
  \caption{Average cycle count per number of evictions in a set.}    
  \label{fig:eviction}
\end{figure}

\begin{figure}[t!]%
  \centering
  \includegraphics[width=\linewidth]{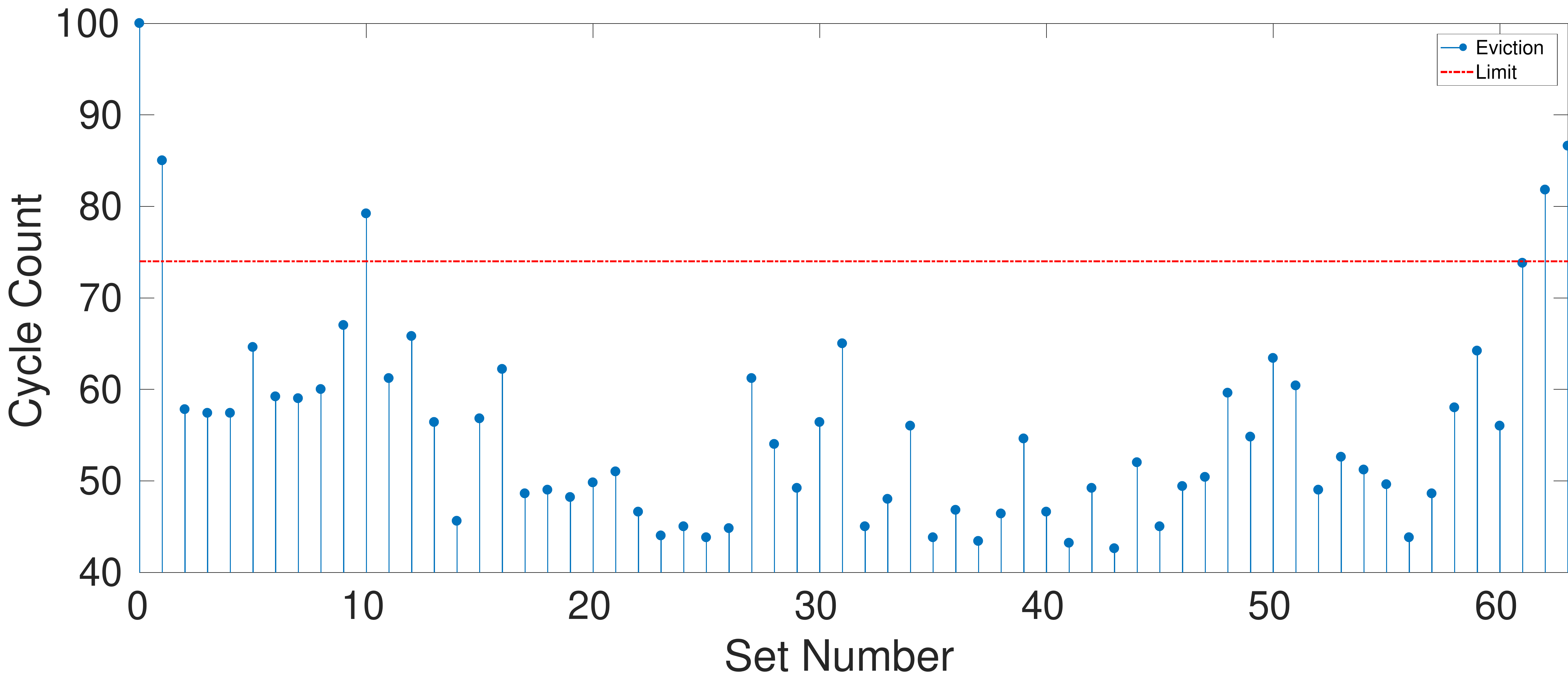}
  \caption{Average cycle count for each set. Variations are due to channel noise: 4 sets are unusable for  attacks.}    
  \label{fig:context_switch__noise}
\end{figure}

%
%\begin{figure}[t]%
%\RawFloats
%\centering
% \begin{minipage}[t]{.48\linewidth}
%  \centering
%  \includegraphics[width=\linewidth]{images/eviction.eps}
%\vspace{-4.5ex}
%  \caption{Average cycle count per number of evictions in a set.}    
%  \label{fig:eviction}
%\end{minipage}
%~
%\begin{minipage}[t]{.48\linewidth}
%\centering
%  \includegraphics[width=\linewidth]{images/context_switch_noise.eps}
%\vspace{-4.5ex}
%  \caption{Average cycle count for each set. Variations are due to channel noise: 4 sets are unusable for  attacks.}    
%  \label{fig:context_switch__noise}
%\end{minipage}
%\vspace{-4ex}
%\end{figure}

\subsection{Testing the Performance of CacheZoom}

Our experimental setup is a Dell Inspiron 5559 laptop with Intel(R) Skylake Core(TM) i7-6500U processor running Ubuntu 14.04.5 LTS and SGX SDK 1.7. Our target processor has 2 hyper-threaded physical cores. Each physical core has 32\,kB of L1D and 32\,kB of L1I local cache memory. The L1 cache, used as our side channel, is 8 way associative and consists of 64 sets. 

Even though Skylake processors use an adaptive LRU cache replacement policy and the adaptive behavior is undocumented~\cite{gruss2016rowhammer}, our results show that we can still use the pointer chasing eviction set technique~\cite{liu2015last} to detect memory accesses. In the specific case of our L1D cache, the access time for chasing 8 pointers associated to a specific set is about 40 cycles on average. In order to test the resolution of our side channel, we took an average of 50000 samples of all the sets and varied the number of evictions from 0 to 8. The results can be seen in Figure~\ref{fig:eviction}, where the access time is increased by roughly 5 cycles for every additional eviction. Thus, our results show that our eviction set gives us an accurate measurement on the number of evicted lines from a specific set. 

\begin{figure*}[t!]%
{
	\centering
	\begin{subfigure}{\label{fig:1stsimon}
			\includegraphics[width=.48\linewidth]{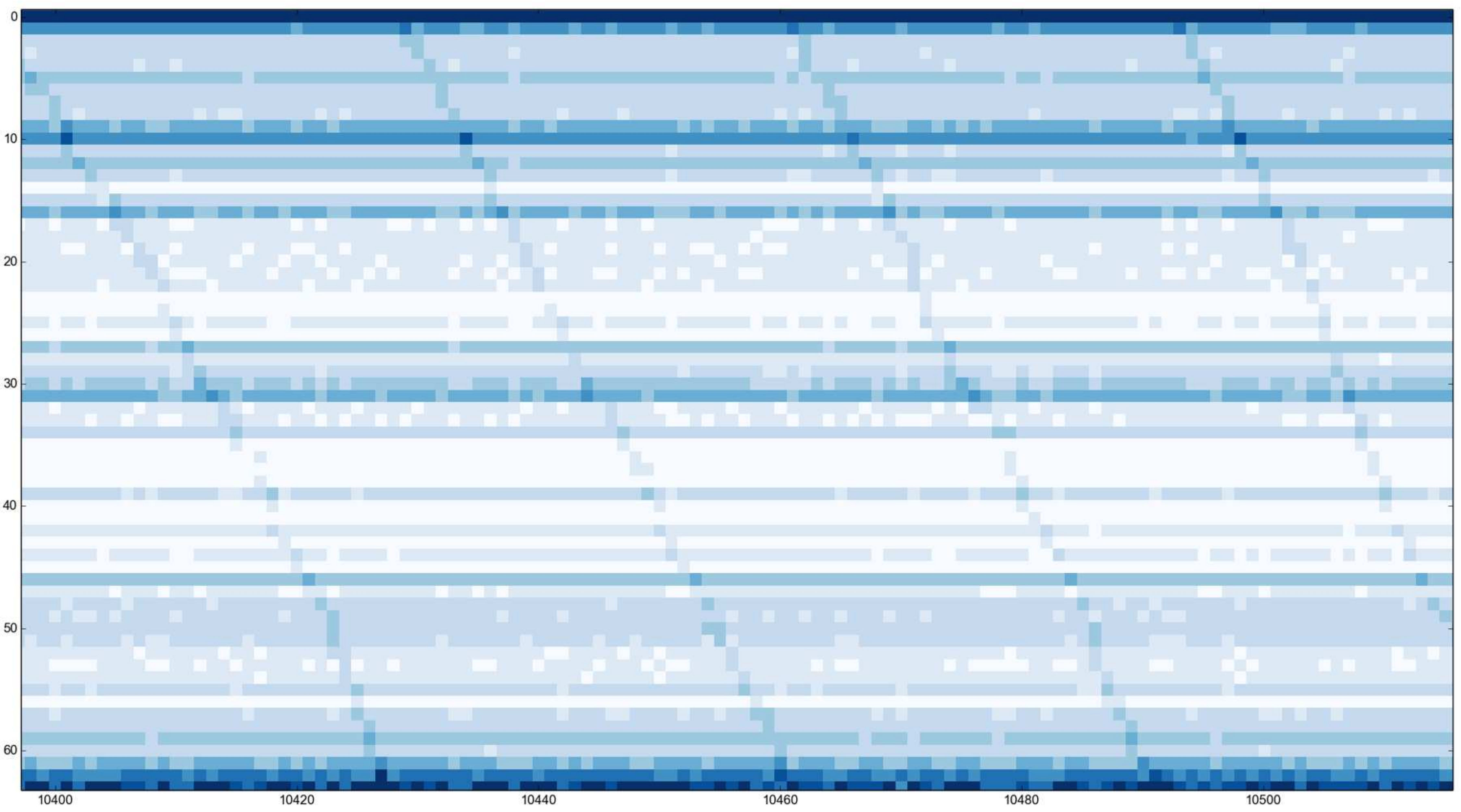}}
	\end{subfigure}%
	\begin{subfigure}{\label{fig:2ndsimon}
			\includegraphics[width=.48\linewidth]{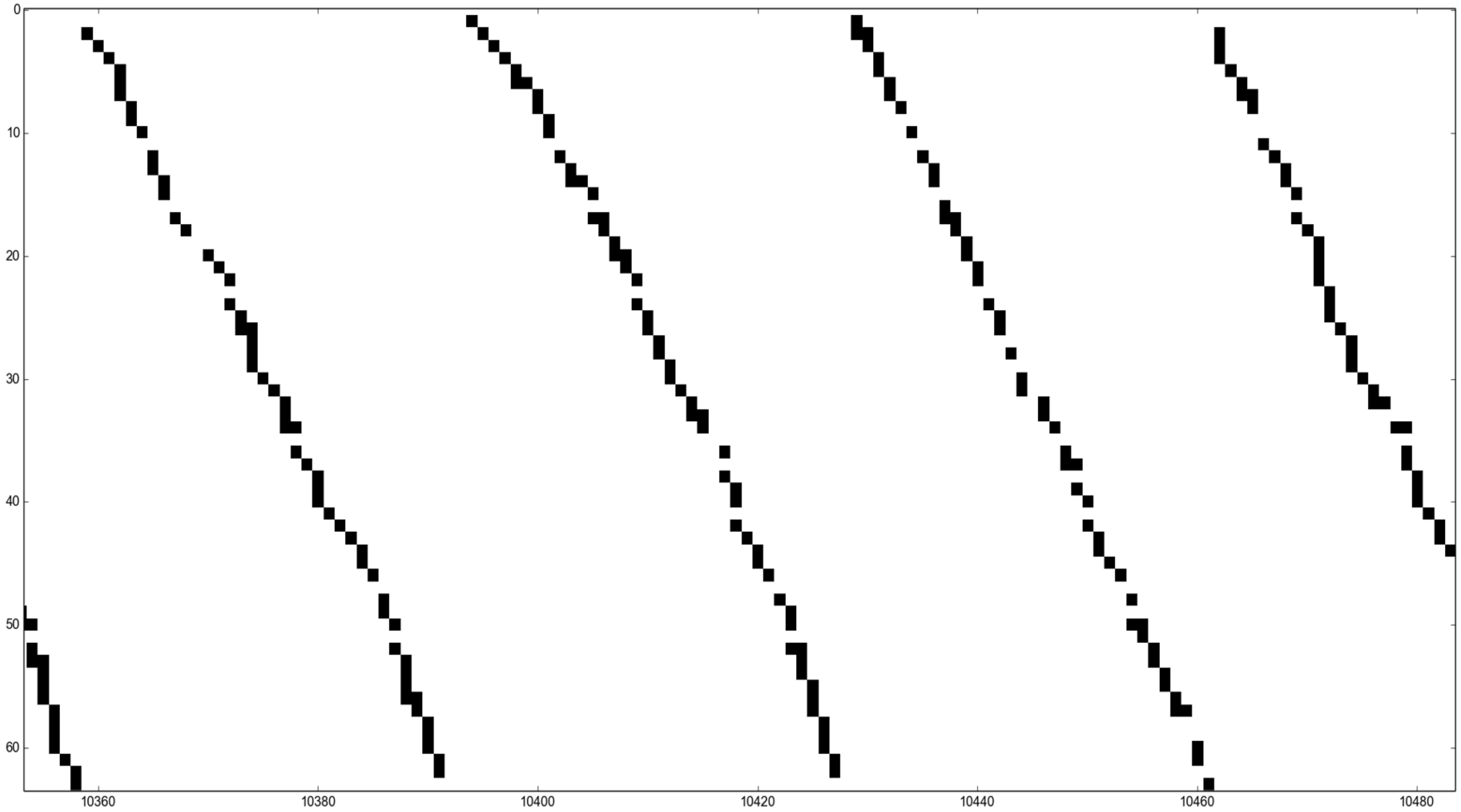}}
	\end{subfigure}%	
     \caption{Cache hit map before (left) and after (right) filtering for context switch noise. Enclave memory access patterns are clearly visible once standard noise from context switch has been eliminated}%
    \label{fig:example}		
}
\end{figure*}

Our isolated CPU core and the L1D eviction set have the minimal possible noise and avoid noises such as CPU frequency, OS and enclave noise; however, the actual noise from the context switch between enclave process and attacker interrupt is mostly unavoidable. The amount of noise that these unwanted memory accesses add to the observation can be measured by running an enclave with an empty loop under our attack measurement. Our results, presented in Figure~\ref{fig:context_switch__noise}, show that every set has a consistent number of evictions. Among the 64 sets, there are only 4 sets that get filled as a side effect of the context switch memory accesses. For the other sets, we observed either 0 or less than 8 unwanted accesses. Due to the consistency of the number of evictions per set, we can conclude that only sets that get completely filled are obscure and do not leak any information, 4 out of 64 sets in our particular case. An example of the applied noise ex-filtration process can be observed in Figure~\ref{fig:example}, in which the enclave process was consecutively accessing different sets. The left hand figure shows the hit access map, without taking into account the appropriate set threshold. The right hand figure shows the access pattern retrieved from the enclave once the context switch noise access has been taking into account and removed.

%\begin{figure}[tb]
%{
%	\centering
%	\begin{subfigure}{\label{fig:1stsimon}
%			\includegraphics[width=5.5cm]{images/pointer_chasing_heatmap.eps}}
%	\end{subfigure}
%	~
%	\begin{subfigure}{\label{fig:2ndsimon}
%			\includegraphics[width=5.5cm]{images/pointer_chasing_heatmap_filtered.eps}}
%	\end{subfigure}%	
%\vspace{-2.5ex}
%    \caption{Cache hit map before (left) and after (right) filtering for context switch noise. Enclave memory access patterns are clearly visible once standard noise from context switch has been eliminated\vspace{-3ex}}%
%    \label{fig:example}		
%}
%\end{figure}

\section{Attack on AES}\label{sec:aesattack}
The following gives a detailed description of different implementation styles for AES to help the reader understand the attacks that we later perform:

\subsection{Cache Attacks on Different AES Implementations}
AES is a widely used block cipher that supports three key sizes from 128 bit to 256 bits. Our description and attacks focus on the 128-bit key version, AES-128, but most attacks described can be applied to larger-key versions as well. AES is based on 4 main operations: AddRoundKey, SubBytes, ShiftRows and MixColumns. The main source of leakage in AES comes from the state-dependent table look ups for the SubBytes operation. These look-ups result in secret-dependent memory accesses, which can be exploited by cache attacks. 
\begin{description}

\item[S-box:] Software implementations that implement the 4 stages independently base the SubBytes operation in a 256 entry substitution table, each entry being 8 bits long. In this implementation, a total of a 160 accesses are performed to the S-box during a 128-bit AES encryption, 16 accesses per round. We refer to this implementation style as the \emph{S-box} implementation.
\item[4 T-tables:] To achieve a better performance, some implementations combine the MixColumns and SubBytes in a single table lookup. At the cost of bigger pre-computed tables (and therefore, more memory usage) the encryption time can be significantly reduced. The most common type uses 4 T-tables: 256 entry substitution tables, each entry being 32 bits long. The entries of the four T-tables are the same bytes but rotated by 1, 2 and 3 positions, depending on the position of the input byte in the column of the AES state. We refer to this style as \emph{T-table implementation}. We refer to this as the 4 T-table implementation.
\item[Large T-table] Aiming at improving the memory usage of T-table based implementations, some designs utilize a single 256 entries T-table,  where each entry is 64 bits long. Each entry contains two copies of the 32 bit values typically observed with regular size T-tables. This design reads each entry \emph{with a different byte offset}, such that the values from the 4 T-tables can be read from a single bigger T-table. The performance of the implementation is comparable, but requires efficient non word-aligned memory accesses. We refer to this as the Large T-table implementation.
\end{description}

Depending on the implementation style, implementations can be more susceptible to cache attacks or less. The resolution an attacker gets depends on the cache line size, which is 64 bytes on our target architecture. 
For the \textbf{S-box} implementation, the S-box occupies a total of 4 cache lines (256 bytes). That is, an attacker able to learn for each observed access to a table entry at most two bits. Attacks relying on probabilistic observations of the S-box entries not being accessed during an entire encryption~\cite{waitaminute} would observe such a case with a probability of $1.02\cdot 10^{-20}$, making a micro-architectural attack nearly infeasible.
For a \textbf{4 T-tables} implementation, each of the T-tables gets 40 accesses per encryption, 4 per round, and occupies 16 cache lines. Therefore, the probability of a table entry not being accessed in an entire encryption is $8\%$, a fact that was exploited in~\cite{waitaminute} to recover the full key. In particular, all these works target either the first or the last round to avoid the MixColumns operation. In the first round, the intermediate state before the MixColumns operation is given by $s^0_i=T_i[p_i \oplus k^0_i]$, where $p_i$ and $k^0_i$ are the plaintext and first round key bytes $i$, $T_i$ is the table utilization corresponding to byte $i$ and $s^0_i$ is the intermediate state before the MixColumns operation in the first round. We see that, if the attacker knows the table entry being utilized $x_i$ and the plaintext he can derive equations in the form $x_i$=$p_i \oplus k^0_i$ to recover the key. A similar approach can be utilized to mount an attack in the last round where the output is in the form $c_i=T_i[s^9_i] \oplus k^{10}_i$.
The maximum an attacker can learn, however, is 4 bit per lookup, if each lookup can be observed separately. The scenario for attacks looking at accesses to a single cache line for an entire encryption learn a lot less, hence need significantly more measurements. 

For a Large T-table implementation, the T-table occupies 32 cache lines, and the probability of not accessing an entry is reduced to $0.6\%$. This, although not exploited in a realistic attack, could lead to key recovery with sufficiently many measurements. An adversary observing each memory access separately, however, can learn 5 bits per access, as each cache line contains only 8 of the larger entries. 

Note that an attacker that gets to observe every single access of the aforementioned AES implementations would succeed to recover the key with significantly fewer traces, as she gets to know the entry accessed at every point in the execution. This scenario was analyzed in~\cite{efficient} with simulated cache traces. Their work focuses on recovering the key based on observations made in the first and second AES rounds establishing relations between the first and second round keys. As a result, they succeed on recovering an AES key from a 4 T-table implementation with as few as six observed encryptions in a noise free environment.

\begin{figure*}[t!]
  \centering
\includegraphics[width=.95\linewidth]{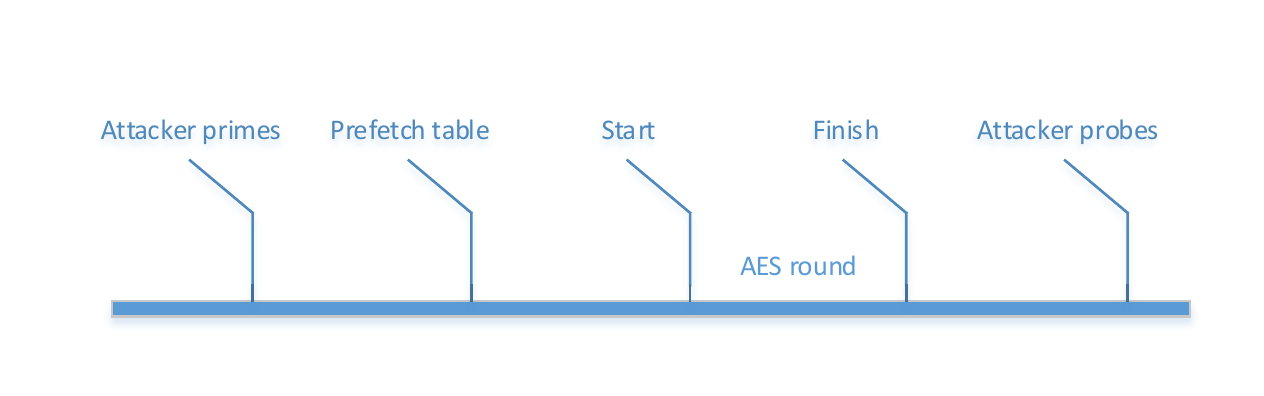}

  \caption{Prefetching and the timeline effect for a regular \PnP\ attack.} %
	 \label{fig:prefetch}
\end{figure*}

\subsection{Non-vulnerable AES Implementations}
There are further efficient implementations of AES that are not automatically susceptible to cache attacks, as they avoid secret-dependent memory accesses. These implementation styles include bit-sliced implementations~\cite{Matsui2007}, implementations using vector instructions~\cite{Hamburg2009}, constant memory access implementations and implementations using AES instruction set extensions on modern Intel CPUs~\cite{aesni}.
However, they all come with their separate drawbacks. The bit-sliced implementations need data to be reformatted before and after encryption and usually show good performance only if data is processed in large chunks~\cite{Bernstein2008}. Constant memory access implementations also suffer from performance as the number of memory accesses during an encryption significantly increases. While hardware support like AES-NI combines absence of leakage with highest performance, it is only an option if implemented and if the hardware can be trusted~\cite{cryptoeprint:2011:428}, and further might be disabled in BIOS configuration options.

%\begin{figure}[t!]
%  \centering
%  \begin{center}
%    \makebox[\textwidth]{\includegraphics[width=.8\linewidth]{images/prefetch.pdf}}
%  \end{center}
%\vspace{-8.5ex}
%  \caption{Prefetching and the timeline effect for a regular \PnP\ attack.\vspace{-3.5ex}} %
%	 \label{fig:prefetch}
%\end{figure}

\subsection{Cache Prefetching as a Countermeasure}
In response to cache attacks in general and AES attacks in particular, several cryptographic library designers implement cache prefetching approaches, which just load the key dependent data or instructions to the cache prior to their possible utilization. In the case of AES, this simply means loading all the substitution tables to the cache, either once during the encryption (at the beginning) or before each round of AES. Prefetching takes advantage of the low temporal resolution that an attacker obtains when performing a regular non-OS controlled attack, as it assumes that an attacker cannot probe faster than the prefetching. Translated to AES, prefetching assumes that a cache attack does not have enough temporal granularity to determine which positions in the substitution table have been used if they are prefetched, e.g., at the beginning of each round.

An example of the implications that such a countermeasure will have on a typical cache attack can be observed in Figure~\ref{fig:prefetch}. The \PnP\ process cannot be executed within the execution of a single AES round. Thanks to prefetching, the attacker is only able to see cache hits on all the Table entries. We analyze whether those countermeasures, implemented in many cryptographic libraries, resist the scenario in which an attacker fully controls the OS and can interrupt the AES process after every small number of accesses. As it was explained in Section~\ref{sec:background}, attacking SGX gives a malicious OS advarsary almost full temporal resolution, which can reverse the effect of prefetching mechanisms.

\section{CacheZooming SGX-based AES}
We use CacheZoom to retrieve secret keys of different implementations of AES running inside an enclave. As mentioned in section \ref{AttackerCapabilities}, we assume no knowledge of the encryption key, but to have access to the enclave binary, and thus to the offset of the substitution tables. We further assume the enclave is performing encryptions over a set of known plaintext bytes or ciphertext bytes.

\begin{figure*}[t!]%
  \centering
  \includegraphics[width=\linewidth]{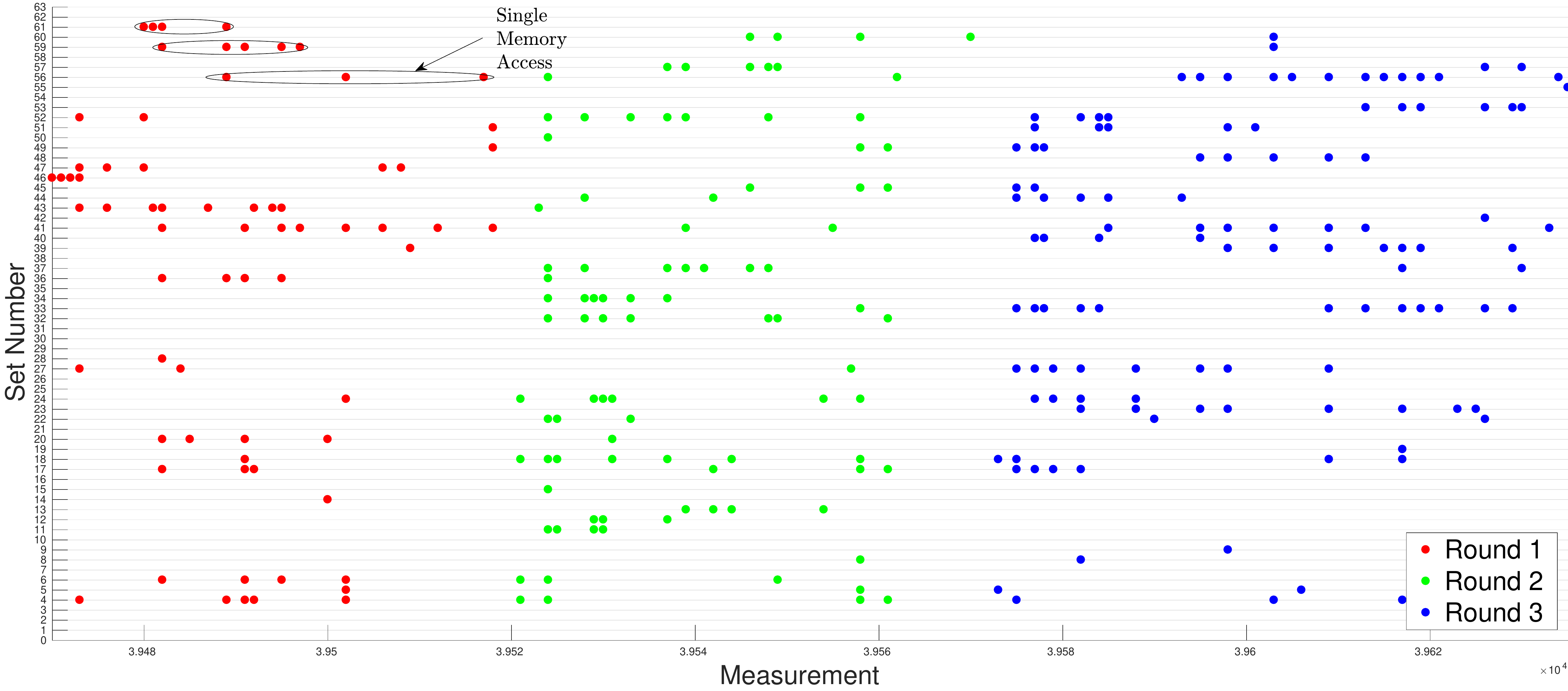}
  \caption{Memory footprint of the AES execution inside enclave.}
	 \label{fig:AESRunFootPrint}
\end{figure*}

\subsection{T-table Implementations}
Our first attacks target the T-table implementations.
To recover the AES key from as few traces as possible, we recover the memory access pattern of the first 2 rounds of the AES function. A perfect single trace first round attack reveals at most the least significant 4 and 5 bits of each key byte in 4 T-table (16 entries/cache line) and Large T-table implementations (8 entries/cache line) respectively. As we want to retrieve the key with the minimal number of traces, we also retrieve the information from the accesses in the second round and use the relation between the first and second round key. In particular, we utilize the relations described in~\cite{efficient}, who utilized simulated data to demonstrate the effectiveness of their AES key recovery algorithm. 

%\begin{figure}[t!]
%  \centering
%\includegraphics[width=\linewidth]{images/sample_run_data.eps}  
%  \vspace{-4ex}
%  \caption{Memory footprint of the AES execution inside enclave.\vspace{-3.5ex}}
%	 \label{fig:AESRunFootPrint}
%\end{figure}

In our specific practical attack, we face three problems: \textbf{(1)} Even in our high resolution attack, we have noise that adds false positives and negatives to our observed memory access patterns. \textbf{(2)} Our experiments show that the out-of-order execution and parallel processing of memory accesses does not allow for a full serialization of the observed memory accesses. \textbf{(3)} Separating memory accesses belonging to different rounds can be challenging. These first two facts can be observed in Figure~\ref{fig:AESRunFootPrint}, which shows 16 memory accesses to each round of a 4 T-table (4 access per table) AES. Due to our high resolution channel and the out-of-order execution of instructions, we observe that we interrupt the out-of-order execution pipeline while a future memory access is being fetched. Thus, interrupting the processor and evicting the entire L1D cache on each measurement forces the processor to repeatedly load the cache line memory until the target read instruction execution completes. 
Hence, attributing observed accesses to actual memory accesses in the code is not trivial.
Although this behavior adds some confusion, we show that observed accesses still have minimal order that we can take into account. As for the third fact, it involves thorough visual inspection of the collected trace. In particular, we realized that every round start involves the utilization of a substantially higher number of sets than the rest, also observable in Figure~\ref{fig:AESRunFootPrint}.

\begin{figure}[t!]%
  \centering
\includegraphics[width=1\linewidth]{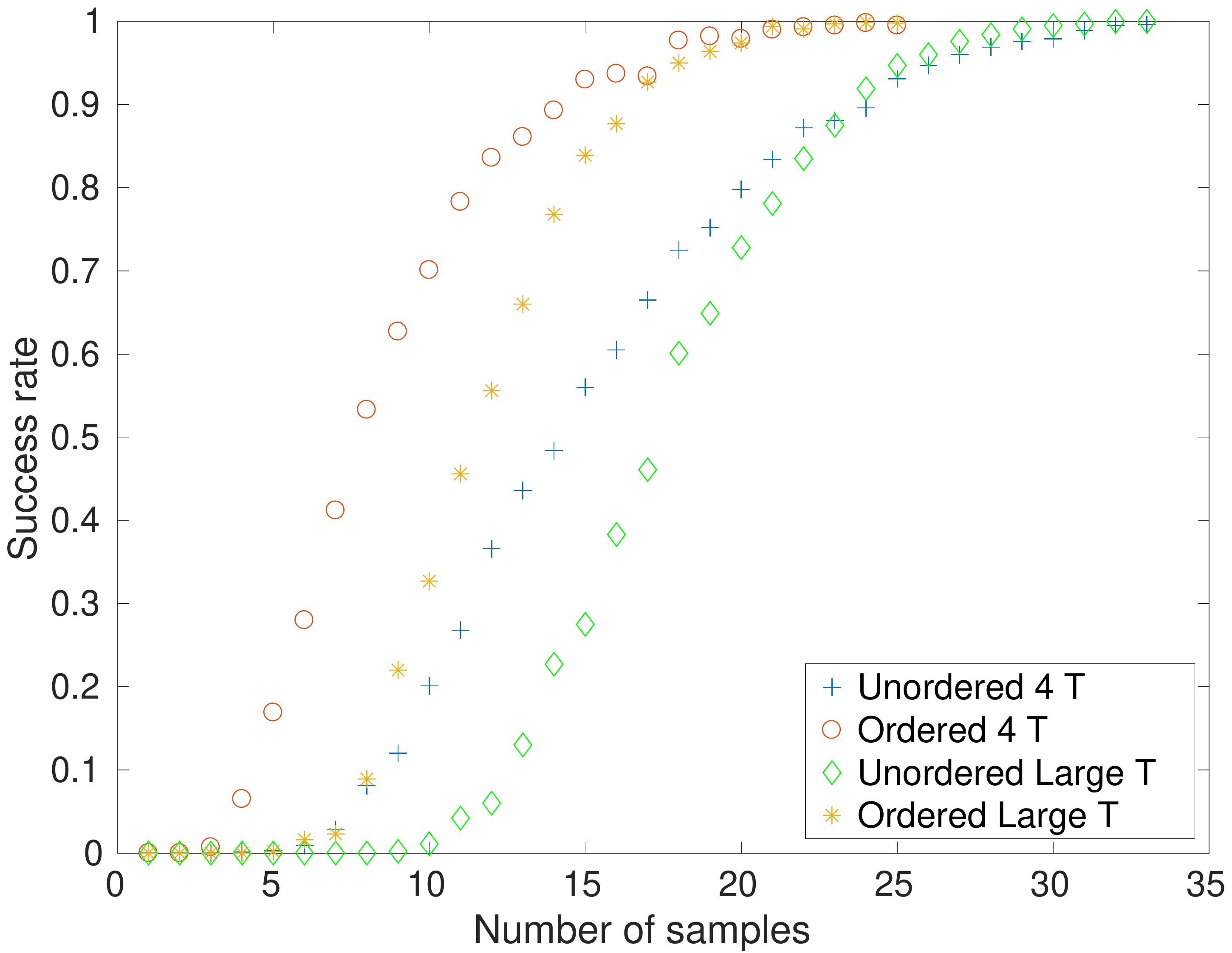}
  \caption{Key recovery success rate.}%
	\label{fig:ttableKeyRecoverySuccessRate}
\end{figure}

\

\begin{table}[tbh]
\centering
 \caption{Statistics on recovered memory accesses for T-table implementations.}%
 \label{tab:statistics}
\begin{tabular}[t]{ |c|c|c| } 
\hline
Implementation & 4 T-table & Large T-table  \\
\hline
True Positive  & 55\% & 75\% \\
False Positive & 44\% & 24\% \\
False Negative & 56\% & 12\% \\
Ordered        & 77\% & 67\% \\
\hline
\end{tabular}
\end{table}

\begin{figure*}[t!]%
  \centering
\includegraphics[width=\linewidth]{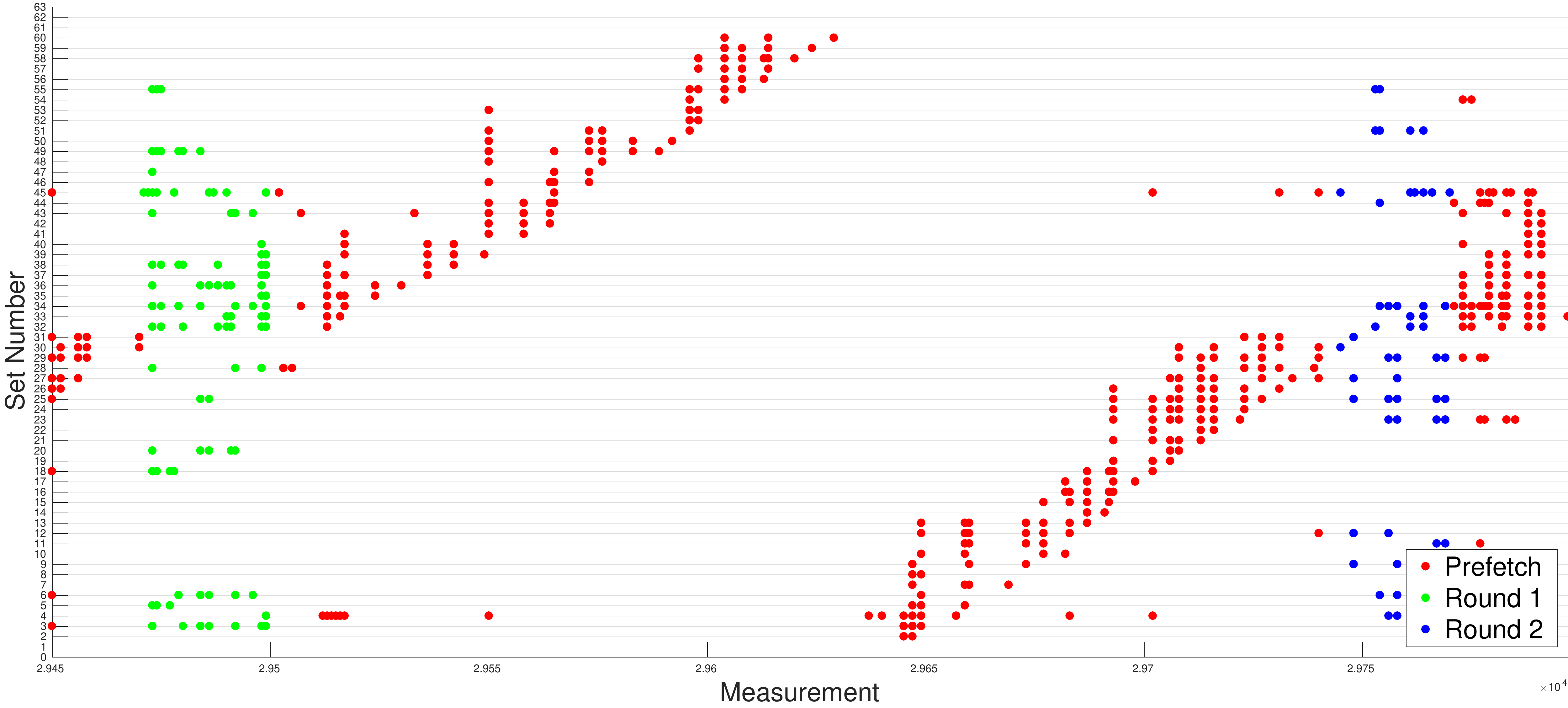}
  \caption{Memory footprint of the AES execution inside an enclave with prefetch countermeasure. The prefetch is clearly distinguishable and helps to identify the start of each round. Further, it also highlights out-of-order execution and in-order completion. } 
	\label{fig:AESRunFootPrintWithPrefetch}
\end{figure*}

In the first implementation of our key recovery algorithm, we just use the set access information without taking into account the ordering of our observed accesses. Recall that we have access to the binary executed by the enclave, and thus, we can map each set number to its corresponding T-table entry. This means that all our accesses can be grouped on a T-table basis. Duplicated accesses to a set within a round are not separated and are considered part of the same access. 
After applying this filter to the first and second round traces, we apply the key recovery algorithm, as explained in~\cite{efficient}. The accuracy of our measurements with respect to the aforementioned issues can be seen in Table~\ref{tab:statistics}. For the 4 T-table implementation, 55\% of the accesses correspond to true accesses (77\% of them were ordered), 44\% of them were noisy accesses and 56\% of the true accesses were missed. For the single Large T-table implementation, 75\% of the T-table accesses corresponded to true accesses (67\% ordered), 24\% were noisy accesses and 12\% of the true accesses were missed. The quality of the data is worse in the 4 T-table case because they occupy larger number of sets and thus include more noisy lines, as explained in Figure~\ref{fig:context_switch__noise}.

With these statistics and after applying our key recovery algorithms with varying number of traces we obtained the results presented in Figure~\ref{fig:ttableKeyRecoverySuccessRate}. If we do not consider the order in our experiments, we need roughly 20 traces (crosses and diamonds) to get the entire correct key with 90\% probability in both the 4 T-table and single T-table implementations. 

%
%\begin{figure}[tb]
%\CenterFloatBoxes
%\begin{floatrow}
%\ffigbox{%
% \includegraphics[width=1\linewidth]{images/ttable_average_success.eps}
%}{%
%  \vspace{-4ex}
%  \caption{Key recovery success rate.  \vspace{-3.5ex}}%
%	\label{fig:ttableKeyRecoverySuccessRate}
%}
%
%\capbtabbox{%
%  \begin{tabular}[t]{ |c|c|c| } 
%	
%\hline
%Implementation & 4 T-table & Large T-table  \\
%\hline
%True Positive  & 55\% & 75\% \\
%False Positive & 44\% & 24\% \\
%False Negative & 56\% & 12\% \\
%Ordered        & 77\% & 67\% \\
%\hline
%\end{tabular}
%}{%	
%  \vspace{-1ex}
%  \caption{Statistics on recovered memory accesses for T-table implementations.}%
%	\label{tab:statistics}
%}
%\end{floatrow}
%\end{figure}

To further improve our results, we attempt to utilize the partial order of the observed accesses. We obtain the average position for all the observed accesses  to a set within one round. These positions are, on average, close to the order in which sets were accessed. The observed order is then mapped to the order in which each T-table should have been utilized. Since this information is not very reliable, we apply a score and make sure misorderings are not automatically discarded. After applying this method, the result for our key recovery algorithm can be observed again in Figure~\ref{fig:ttableKeyRecoverySuccessRate}, for which we needed around 15 traces for the 4 T-table implementation (represented with stars) and 12 traces for the single Large T-table implementation (represented circles) to get the key with 90\% probability. Thus, we can conclude that using the approximate order helped us to recover the key with fewer traces.

\subsubsection{Cache Prefetching\textsf{,}} as explained in Section~\ref{sec:aesattack}, is implemented to prevent passive attackers from recovering AES keys. CacheZoom, in theory, should bypass such a countermeasure by being able to prime the cache after the T-tables are prefetched. The observation of a trace when cache prefetching is implemented before every round can be observed in Figure~\ref{fig:AESRunFootPrintWithPrefetch}. We can see how cache prefetching is far from preventing us to recover the necessary measurements. In fact, it eases the realization of our attack, as we now can clearly distinguish accesses belonging to different rounds, allowing for further automation of our key recovery step. Thus, CacheZoom not only bypasses but further benefits from mechanisms that mitigated previous cache attacks.

%
%
%\begin{figure}[t!]%
%  \centering
%\includegraphics[width=\linewidth]{images/sample_run_data_prefetch.eps}
%  \vspace{-4ex}
%  \caption{Memory footprint of the AES execution inside an enclave with prefetch countermeasure. The prefetch is clearly distinguishable and helps to identify the start of each round. Further, it also highlights out-of-order execution and in-order completion.  \vspace{-3.5ex}} 
%	\label{fig:AESRunFootPrintWithPrefetch}
%\end{figure}

\subsection{S-Box Implementation}

S-box implementation is seen as a remedy to cache attacks, as all S-box accesses use only a very small number of cache lines (typically 4). With 160 S-Box accesses per encryption, each line is loaded with a very high likelihood and thus prevents low resolution attackers from gaining information. Adding a prefetch for each round does not introduce much overhead and also prevents previous attacks that attempted interrupting the execution~\cite{BriongosMRM16,gullasch2011cache}. However, CacheZoom can easily distinguish S-box accesses during the rounds, but due to the out-of order execution, it is not possible to distinguish accesses for different byte positions in a reliable manner. However, one distinguishable feature is the number of accesses each set sees during a round. We hypothesize that the number of observed accesses correlates with the number of S-box lookups to that cache line. If so, a classic DPA correlating the observed accesses to the predicted accesses caused by one state byte should recover the key byte. Hence we followed a classic DPA-like attack on the last round, assuming known ciphertexts. 

The model used is rather simple: for each key byte $k$, the accessed cache set during the last round for a given ciphertext byte $c$ is simply given as $set = S^{-1}(x\oplus k)\gg6$, i.e. the two MSBs of the corresponding state byte before the last SubBytes operation. The access profile for a state byte position under an assumed key $k$ and given ciphertext bytes can be represented by a matrix $A$ where each row corresponds to a known ciphertext and each column indicates whether that ciphertext resulted in an access to the cache line with the same column index. Hence, each row has four entries, one per cache line, where the cache line with an access is set to one, and the other three columns are set to zero (since that state byte did not cause an access). 
Our leakage is given as a matrix $L$, where each row corresponds to a known ciphertext and each column to the number of observed accesses to one of the 4 cache lines. 
A correlation attack can then be performed by computing the correlation between $A$ and $L$, where $A$ is a function of the key hypothesis. We used synthetic, noise-free simulation data for the last AES round to validate our approach, where accesses for 16 bytes are accumulated over 4 cache lines for numerous ciphertexts under a set key. The synthetic data shows a best expectable correlation of about .25 between noise-free cumulative accesses $L$ and the correct accesses for a single key byte $A$. As little as 100 observations yield a first-order success rate of 93\%.
%\begin{figure}[t]%
%\RawFloats
%\centering
%\begin{minipage}[t!]{.48\linewidth}%
%  \centering
%     \includegraphics[width=\linewidth]{images/CorrelationVSkeyPosition.eps}
%\vspace{-4ex}
%    \caption{Correlation between observed and expected accesses caused by one byte position. Leakage is stronger for later bytes. Correlation of observed (blue) vs.\ relative accesses (amber).}
%		\label{fig:CorrelationVSkeyPosition}
%\end{minipage}
%~
% \begin{minipage}[t!]{.48\linewidth}%
%  \centering
%		\includegraphics[width=\linewidth]{images/GoodBadCorr.eps}	
%\vspace{-4ex}
%    \caption{Correlation of key values for the best ($k_{15}$, amber) and worst ($k_0$, blue) key bytes with 1500 traces. The guess with the highest correlation (o) and the correct key (x) match only for $k_{15}$.}
%			\label{fig:GoodBadCorr}
%\end{minipage}
%\vspace{-3.5ex}
%\end{figure}

\begin{figure}[t!]%
  \centering
     \includegraphics[width=\linewidth]{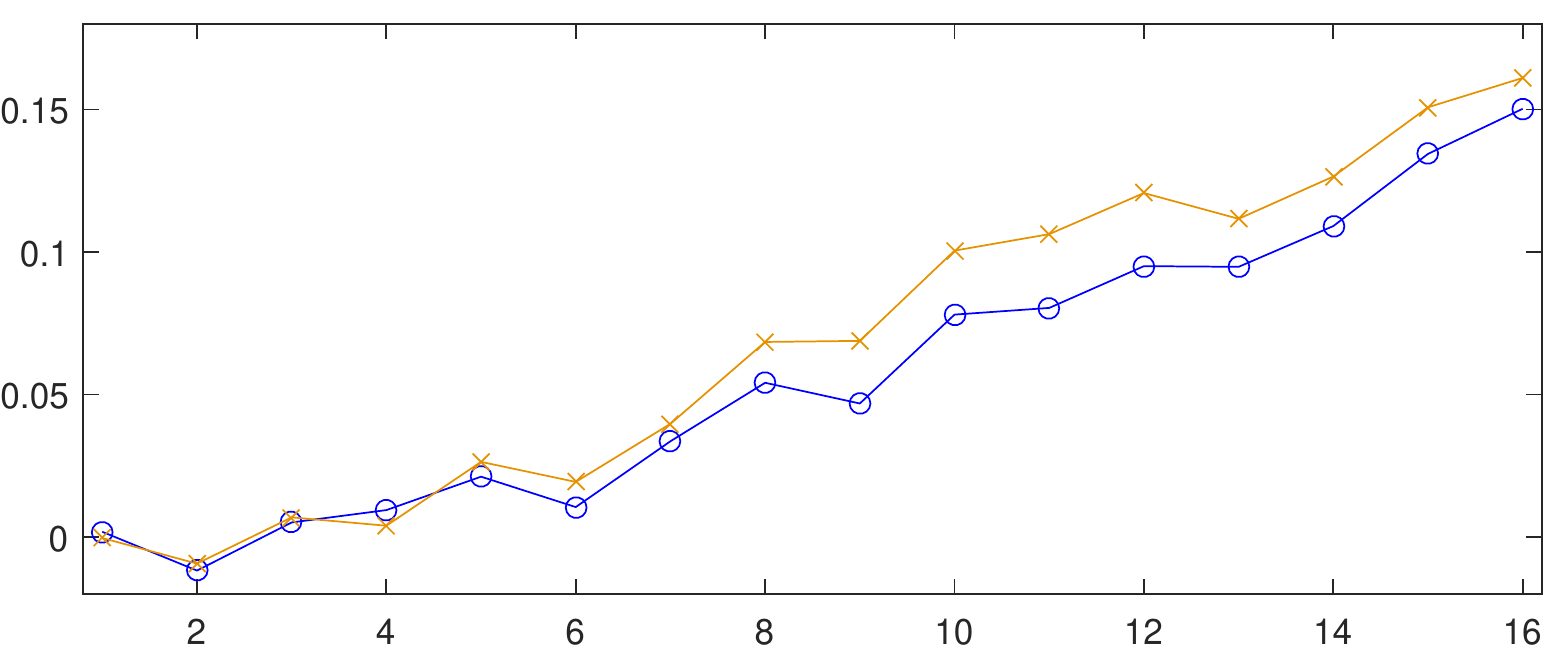}
    \caption{Correlation between observed and expected accesses caused by one byte position. Leakage is stronger for later bytes. Correlation of observed (blue) vs.\ relative accesses (amber).}
		\label{fig:CorrelationVSkeyPosition}
\end{figure}

\begin{figure}[t!]%
  \centering
	\includegraphics[width=\linewidth]{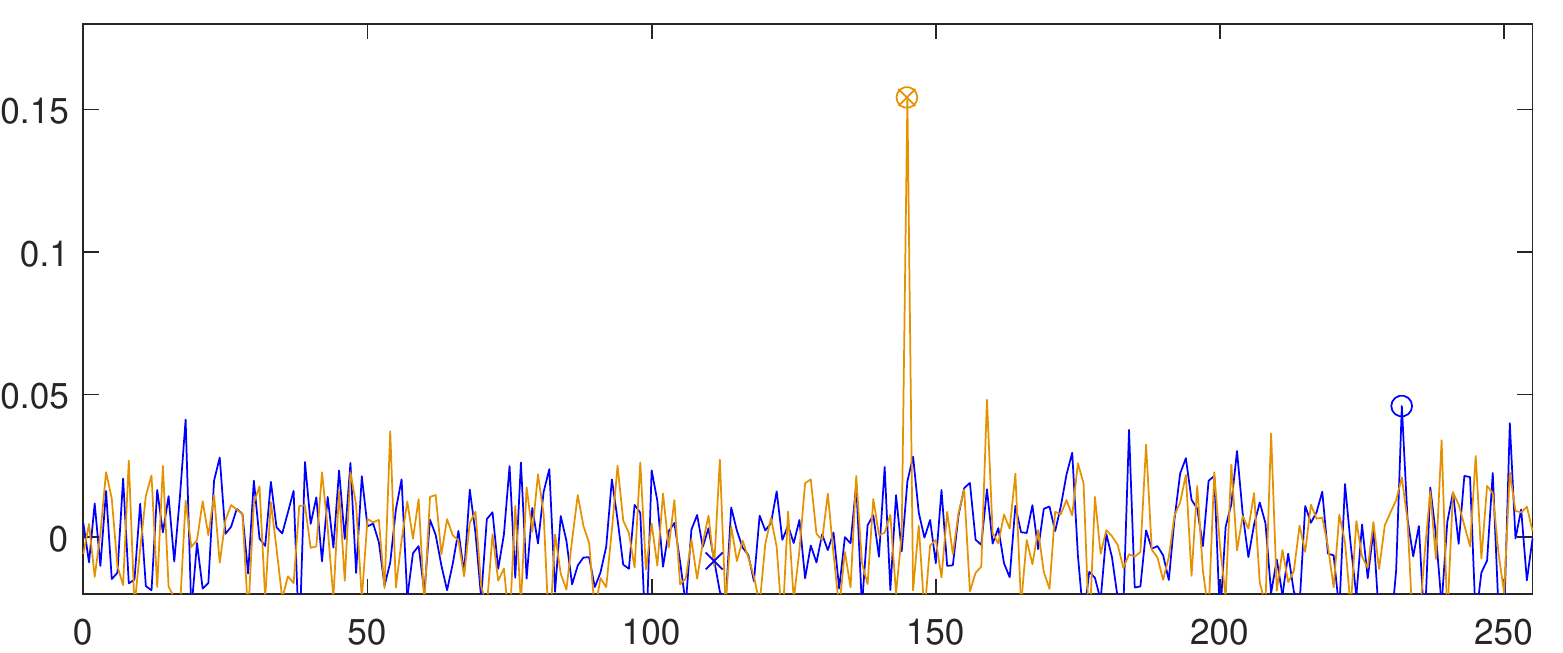}	
    \caption{Correlation of key values for the best ($k_{15}$, amber) and worst ($k_0$, blue) key bytes with 1500 traces. The guess with the highest correlation (o) and the correct key (x) match only for $k_{15}$.}
	\label{fig:GoodBadCorr}
\end{figure}

Next, we gathered hundreds of measurements using CacheZoom. Note that due to a lack of alignment, the collection of a large number of observations and the extraction of the last round information still requires manual intervention. When performing the key recovery attack, even 200 observations yielded 4-5 key bytes correctly. However, the first-order success rate only increases very slowly with further measurements. We further observed that \textbf{(1)} more traces always recover later key bytes first and \textbf{(2)} key ranks for earlier lookups are often very low, i.e. the correct key does not even yield a high correlation.
To analyze this behavior, we simply correlated the expected leakage $A$ for each byte position to the observed leakage $L$. The result is shown in Figure~\ref{fig:CorrelationVSkeyPosition}. It can be observed that the correlation for the later key bytes is much stronger than for the earlier key bytes. This explains why later key bytes are much easier to recover. The plot also shows a comparison of using the absolute number of observed accesses (ranging between 10 and 80 observed accesses per round, blue) an the relative number of accesses per cache set (amber) after removing outliers.

Results for the best and the worst key guess are shown in Figure~\ref{fig:GoodBadCorr}. For $k_{15}$ (amber), the correlation for the correct key guess is clearly distinguishable. For $k_0$ however, the correct key guess does not show any correlation with the used 1500 observations. In summary, 500 traces are sufficient to recover 64 key bits, while 1500 recover 80 key bits reliably. While full key recovery will be challenging, recovering 12 out of 16 key bytes is easily possible with thousands of observations. The remaining key bytes can either be brute-forced or can be recovered by exploiting leakage from the second last round.

Next, we explain the reason why we believe bytes processed first are harder to recover. 
The Intel core i7 uses deep pipelines and speculative out-of-order execution. Up to six micro-instructions can be dispatched per clock cycle, and several instructions can also complete per cycle. 
As a result, getting order information for the accesses is difficult, especially if 16 subsequent S-box reads are spread over only 4 cache lines. While execution is out-of-order, each instruction and its completion state are tracked in the CPU's reorder buffer (ROB). Instruction results only affect the system state once they are completed \emph{and} have reached the top of the ROB. That is, micro-ops \emph{retire} in-order, even though they \emph{execute} out-of-order.
The result of micro-ops that have completed 
hence do not immediately affect the system. In our case, if the previous load has not yet been serviced, the subsequent completed accesses cannot retire and affect the system until the unserviced load is also completed. 

\begin{figure*}[t!]%
\begin{center}
    \captionof{table}{Vulnerable implementations in popular current cryptographic libraries. These implementations can be configured through compile/runtime settings.} \label{tab:libraries} 
    \begin{tabular}{ | l | p{11cm} |}
    \hline
    \textbf{Library}  & \textbf{Vulnerable Implementations} \\ \hline
    OpenSSL 1.1.0f& \texttt{aes\_core.c} T-table, \texttt{aes\_x86core.c} Large T-table, S-box and prefetching configurable through \scriptsize{AES\_COMPACT\_IN\_INNER\_ROUNDS, AES\_COMPACT\_IN\_OUTER\_ROUNDS}\\\hline
    WolfCrypt 3.11.0& \texttt{aes.c} T-Table with prefetching before the first round. \\\hline
    Mozilla NSS 3.30.2 & \texttt{rijndael.c} T-Table and S-box configurable through \scriptsize{RIJNDAEL\_GENERATE VALUES\_MACRO}\\\hline
    Nettle 3.3& \texttt{aes-encrypt-internal.asm} T-table. \\\hline
    Libtomcrypt 1.17& \texttt{aes.c} T-table. \\\hline
    Libgcrypt 1.7.7 & \texttt{rijndael.c} T-table, S-box for the last round with prefetching. \\\hline
    MbedTLS 2.4.2& \texttt{aes.c} T-table, S-box for the last round. \\\hline
    \end{tabular}
\end{center}
\end{figure*}

Every context switch out of an enclave requires the CPU to flush the out-of order execution pipeline of the CPU~\cite{costanintel}. Hence CacheZoom's interrupt causes a pipeline flush in the CPU, all micro-ops on the ROB that are not at the top and completed will be discarded. Since our scheduler switches tasks very frequently, many loads cannot retire and thus the same load operation has to be serviced repeatedly. This explains why we see between 9 and 90 accesses to the S-box cache lines although there are only 16 different loads to 4 different cache lines. The loads for the first S-box are, however, the least affected by preceding loads. Hence, they are the most likely to complete and retire from the ROB after a single cache access. Later accesses are increasingly likely to be serviced more than once, as their completion and retirement is dependent on preceding loads. Since our leakage model assumes such behavior (in fact, we assume one cache access per load), the model becomes increasingly accurate for later accesses.

\section{Conclusion}
This work presented CacheZoom, a new tool to analyze memory accesses of SGX enclaves. To gain maximal resolution, CacheZoom combines a L1 cache \PnP\ attack with OS modifications that greatly enhance the time resolution. SGX makes this scenario realistic, as both a modified OS and knowledge of the unencrypted binary are realistic for enclaves. We demonstrate that CacheZoom can be used to recover key bits from all major software AES implementations, including ones that use prefetches for each round as a cache-attack countermeasure. Furthermore, keys can be recovered with as few as 10 observations for T-table based implementations. For the trickier S-box implementation style, 100s of observations reveal sufficient key information to make full key recovery possible. Prefetching is in this scenario beneficial to the adversary, as it helps identifying and separating the accesses for different rounds. A list of libraries that contain vulnerable implementations can be found at table \ref{tab:libraries}.

CacheZoom serves as evidence that security-critical code needs constant execution flows and secret-independent memory accesses.  As SGX's intended use is the protection of sensitive information, enclave developers must thus use the necessary care when developing code and avoid microarchitectural leakages.
For AES specifically, SGX implementations must feature constant memory accesses. Possible implementation styles are thus bit-sliced or vectorized-instruction-based implementations or implementations that access all cache lines for each look-up. 

\smallskip
\noindent\textbf{Acknowledgments} This work is supported by the National Science Foundation, under the grant CNS-1618837. 

\emph{CacheZoom} source repository and data sets are available at \url{https://github.com/vernamlab/CacheZoom}.

%\bibliographystyle{splncs03}
%\bibliography{reference}

{\footnotesize \bibliographystyle{acm}
\bibliography{reference}
}

%\appendix
%\section{Vulnerable AES Implementations}

\end{document}